\documentclass [twocolumn, aps, prl, superscriptaddress, showpacs] {revtex4}
\usepackage{graphicx,latexsym}
\begin{document}
\draft
\title {Quantum Hall effect in dual-gated graphene bilayers with tunable layer density imbalance}

\author {Seyoung Kim}
\author {E. Tutuc}
\affiliation {Microelectronics Research Center, The University of
Texas at Austin, Austin, TX 78758}
\date{\today}
\begin{abstract}
We study the magnetotransport properties of dual-gated graphene
bilayers, in which the total density and layer density imbalance
are independently controlled. As the bilayer is imbalanced we
observe the emergence of a quantum Hall state (QHS) at filling
factor $\nu=0$ evinced by a plateau in the Hall conductivity,
consistent with the opening of a gap between the electron and hole
bands. By varying the layer density imbalance at fixed total
density, we observe a suppression of the QHS at filling factors
$\nu=8$ and $\nu=12$ when the layer densities are balanced, an
observation at variance with theoretical expectations in the
absence of electron-electron interaction and disorder.
\end{abstract}
 \maketitle

Graphene, a layer of carbon atoms in a honeycomb lattice, has
emerged in recent years as a new test-bed for electron physics in
reduced dimensions \cite{novoselov04, novoselov06}. The linear
energy band dispersion, zero energy band-gap, and chiral
quasi-particles set this material apart from conventional
two-dimensional electron systems (2DES) realized in semiconductor
heterostructures. Graphite bilayers, consisting of two closely
coupled graphene layers, are an equally interesting system with
parabolic momentum-energy dispersion and chiral quasi particles
possessing 2$\pi$ Berry phase \cite{novoselov_natphys06}. Most
interesting, in dual-gated graphene bilayers the electron on-site
energy and density in each layer can be independently controlled,
which in turn enables the band-gap energy tuning. Here we
investigate the magnetotransport properties of dual-gated graphene
bilayers with high-k dielectrics, a device geometry that allows an
independent control of the total density and layer density
imbalance. As the bilayer is imbalanced we observe the emergence
of a quantum Hall state (QHS) at filling factor $\nu=0$, evinced
by a plateau in the Hall conductivity, consistent with the opening
of a gap between the electron and hole bands. By varying the layer
density imbalance at fixed total density, we observe an unexpected
suppression of the QHSs at filling factors $\nu=8$ and $\nu=12$
when the layer densities are closely balanced. This observation is
at variance with the single particle picture, strongly suggesting
that electron-electron interaction or disorder play a role in this
system.

Our samples consist of natural graphite mechanically exfoliated
\cite{ngs} on a 300 nm SiO$_2$ dielectric layer, thermally grown
on a highly doped $n$-type Si substrate, with an As doping
concentration of $\sim10^{20}$ cm$^{-3}$. Optical inspection
combined with Raman scattering are used to identify graphene
bilayer flakes for device fabrication. We define six metal
contacts using electron beam (e-beam) lithography followed by 50
nm Ni deposition and lift-off [Fig. 1(a)]. A second e-beam
lithography step followed by O$_2$ plasma etching are used to
pattern a Hall bar on the graphene bilayer flake. A key issue for
graphene gated devices is the deposition of a top dielectric
without degrading the carrier mobility. Owing to the chemically
inertness of graphite, attempts to grow high-$k$ dielectrics, such
as Al$_2$O$_3$ or HfO$_2$, by atomic layer deposition (ALD) on
$\it{clean}$ highly oriented pyrolytic graphite leads to selective
growth on terraces, where broken carbon bonds serve as nucleation
centers. To deposit an Al$_2$O$_3$ top dielectric on our graphene
bilayer samples, we first deposit a $\sim20\AA$ thin Al layer,
which serves as a nucleation layer for the ALD of Al$_2$O$_3$. The
sample is then taken out in air and transferred to an ALD chamber.
Based on X-ray photoelectron spectroscopy and electrical
measurements, the Al layer is fully oxidized thanks to the
presence of residual O$_2$ in the Al evaporation chamber and the
exposure to ambient O$_2$ \cite{dignam}. Next a 15 nm-thick
Al$_2$O$_3$ film is deposited using trimethyl aluminum as the Al
source and H$_2$O as oxidizer. Lastly, a Ni top gate is deposited
using e-beam lithography, metal deposition, and lift off [Fig.
1(a)]. The dielectric deposition technique used here has been
shown to produce top-gated monolayer graphene samples with high,
over 8,000 cm$^2$/Vs, carrier mobility \cite{kim}. The mobility of
the dual-gated bilayer graphene samples investigated here is over
$\simeq~2200$ cm$^2$/Vs \cite{adam}. Longitudinal ($\rho_{xx}$)
and Hall ($\rho_{xy}$) resistivity measurements are performed down
to a temperature of $T=0.3$ K, and using standard low-current,
low-frequency lock-in techniques.

Hall measurements allow us to determine the total carrier density
($n_{tot}$) as a function of $V_{TG}$ and $V_{BG}$, as well as the
corresponding capacitance values. For the sample investigated here
the top- and back-gate capacitances are $C_{TG}=225$
nF$\cdot$cm$^{-2}$ and $C_{BG}=10$ nF$\cdot$cm$^{-2}$,
respectively. A second parameter relevant for graphene bilayers is
the layer density imbalance $\Delta n=(n_B-n_T)/2$, defined in
terms of the difference between the bottom ($n_B$) and top ($n_T$)
layer densities. The use of top- and back- gate allows us to
independently control $n_{tot}$ and $\Delta n$. Up to an additive
constant, $n_{tot}$ and $\Delta n$ are related to $V_{TG}$ and
$V_{BG}$ by $n_{tot}=(C_{BG}\cdot V_{BG}+C_{TG} \cdot V_{TG})/e$,
and $\Delta n=(C_{BG}\cdot V_{BG}-C_{TG} \cdot V_{TG})/2e$; $e$ is
the electron charge. The definition of $\Delta n$ represents the
layer density imbalance in the limit of excellent screening in
each layer, when the top and bottom gates control only the charge
density in the top and bottom layers, respectively; the actual
layer density imbalance is smaller than $\Delta n$ \cite{mccann}.
The layer density imbalance translates into a transverse electric
field on the graphene bilayer, $E=e \Delta n/\varepsilon_0$;
$\varepsilon_0$ is the vacuum dielectric permitivity.

\begin{figure}
\centering
\includegraphics[scale=0.4]{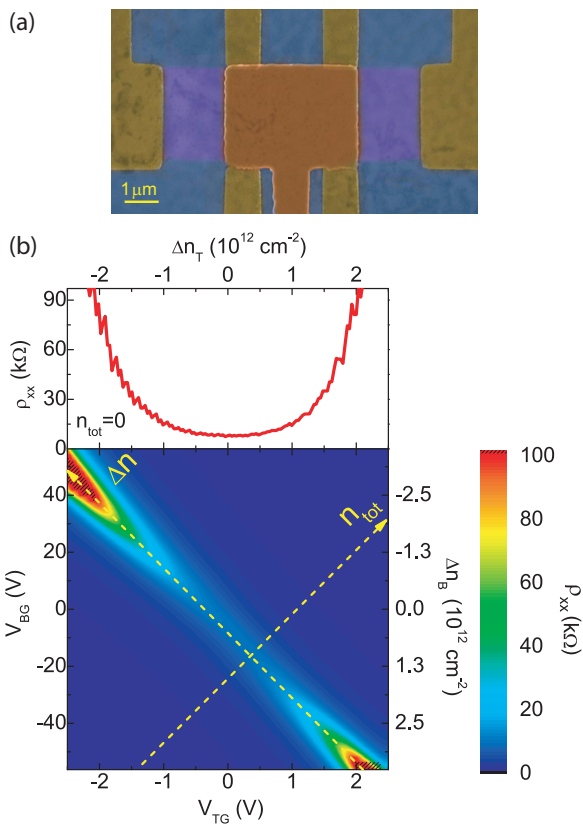}
\caption {\small{(color online) (a) Scanning electron micrograph
showing a dual gated graphene bilayer. The contact and top-gate
terminals have been color coded in yellow and red, respectively.
(b) Contour plot of the bilayer $\rho_{xx}$ measured as a function
of $V_{TG}$ and $V_{BG}$. The right and top axis represent the
corresponding charge density change for the back- and top-gates,
respectively. The top panel shows $\rho_{xx}$ vs $V_{TG}$ measured
at $n_{tot}=0$, which represents the cut along the $\Delta n$
diagonal of the contour plot. These data allow a one-to-one
correspondence between $V_{TG}, V_{BG}$ on one hand, and $n_{tot},
\Delta n$ on the other.}}
\end{figure}

In Fig. 1(b) we show the sample longitudinal resistivity
($\rho_{xx}$) measured as a function of top (V$_{TG}$) and back
($V_{BG}$) gate biases, at a temperature $T=0.3$ K. The diagonals
of constant $C_{BG} \cdot V_{BG} + C_{TG} \cdot  V_{TG}$ represent
the loci of constant $n_{tot}$ and varying $\Delta n$, while
diagonals of constant $C_{BG} \cdot V_{BG}-C_{TG} \cdot V_{TG}$
define the loci of constant $\Delta n$ at varying $n_{tot}$. The
diagonal of $n_{tot}=0$ is defined by the points of maximum
$\rho_{xx}$ measured as a function of $V_{TG}$ at fixed $V_{BG}$
values. In order to determine the V$_{TG}$ and $V_{BG}$ values at
which $n_{tot}=0$ and $\Delta n=0$, we consider $\rho_{xx}$
measured as a function of $V_{TG}$ along the diagonal $n_{tot}=0$
[Fig. 1(b) (top panel)]. As a function of $V_{TG}$, $\rho_{xx}$
displays a minimum and increases for both negative and positive
$V_{TG}$. This trend can be explained by the opening of a band gap
in the graphene bilayer, as a result of different electron on-site
energies on the two layers \cite{mccannprl06,mccann,min}. The
band-gap increases with the transverse electric field and results
in a reduced electrical conductivity \cite{ohta,oostinga}. The
$\rho_{xx}$ dependence on $V_{TG}$ along the diagonal $n_{tot}=0$
allows us to determine the gate biases at which $\Delta n=0$. The
$\rho_{xx}$ minimum on the $n_{tot}=0$ diagonal of Fig. 1(b)(top
panel) defines the $\Delta n=0$ point \cite{error}. Having
established a one-to-one correspondence between $V_{TG}$ and
$V_{BG}$ on one hand, and $n_{tot}$ and $\Delta n$ on the other,
in the reminder of the manuscript we will characterize the bilayer
in terms of $n_{tot}$ and $\Delta n$.

\begin{figure}
\centering
\includegraphics[scale=0.56]{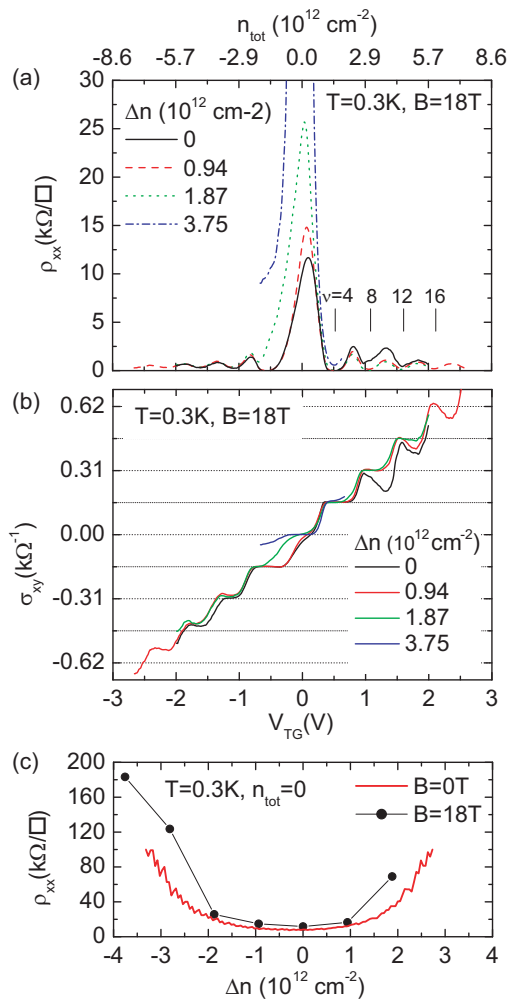}
\caption {\small{(color online) (a) $\rho_{xx}$ vs. $n_{tot}$
measured at $B=18$ T and $T=0.3$ K, for different values of
$\Delta n$. (b) $\sigma_{xy}$ vs. $n_{tot}$ measured at different
$\Delta n$ values, and at $B=18$ T and $T=0.3$ K. The bottom axis
shows the corresponding $V_{TG}$ value measured with respect to
the $n_{tot}$ and $\Delta n$ neutrality point; the $V_{BG}$ values
(not shown) can be inferred using the $C_{TG}$ and $C_{BG}$
values. The horizontal grid lines are spaced by 4$e^2/h$. (c)
$\rho_{xx}$ vs. $\Delta n$ measured at $n_{tot}=0$ and $B=0$
(line), and at $\nu=0$ and $B=18$ T (symbols).}}
\end{figure}

The dual-gated device of Fig. 1(a) allows independent control of
$n_{tot}$ and $\Delta n$. In Fig. 2(a) we show $\rho_{xx}$ vs
$n_{tot}$, measured at {\it fixed} values of $\Delta n$, at a
magnetic field $B=18$ T and at $T=0.3$ K. These data are measured
by simultaneously sweeping $V_{TG}$ and $V_{BG}$, with the sweep
rates adjusted such that $\Delta n$ remains constant. The data
shows QHSs, marked by vanishing $\rho_{xx}$ at integer filling
factors that are multiples of four \cite{filling}. This
observation is explained by the four-fold degeneracy associated
with both spin and valley degrees of freedom of each Landau level
\cite{mccannprl06}. Using the measured $\rho_{xx}$ and
$\rho_{xy}$, we determine the Hall conductivity ($\sigma_{xy}$)
via a tensor inversion,
$\sigma_{xy}=\rho_{xy}/(\rho_{xx}^2+\rho_{xy}^2)$. Figure 2(b)
data shows the Hall conductivity ($\sigma_{xy}$) measured as a
function of $n_{tot}$, at $B=18$ T and $T=0.3$ K, and for
different, fixed values of $\Delta n$. Interestingly, the data of
Fig. 2(a,b) reveals an increasing $\rho_{xx}$ at $n_{tot}=0$ with
increasing $\Delta n$, accompanied by the emergence of a plateau
in the Hall conductivity at $\sigma_{xy}=0$, which in turn
indicates an emerging QHS at $\nu=0$.

\begin{figure}
\centering
\includegraphics[scale=0.4]{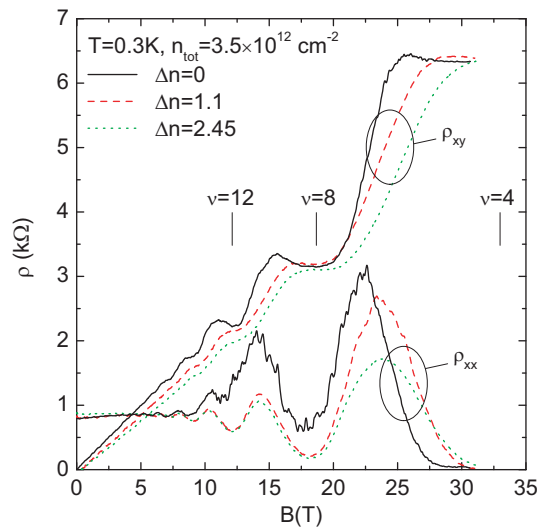}
\caption {\small{(color online) $\rho_{xx}$ and $\rho_{xy}$ vs.
$B$, measured at $T=0.3$ K and at different $\Delta n$ values. The
total density is constant, $n_{tot}=3.5\times10^{12}$ cm$^{-2}$.}}
\end{figure}

The emergence of a QHS at $\nu=0$ in graphene bilayers can be
explained by considering the electron energy levels in this system
with and without an applied magnetic field.
At $E=0$ and $B=0$ the graphene band structure has symmetric,
parabolic electron and hole bands with a zero energy gap, and an
effective mass $m\approx0.054m_0$; $m_0$ is the free electron mass
\cite{mccannprl06}. The electron and hole band are four-fold
degenerate, owing to the spin and valley degrees of freedom. At
finite $E$-field, or $\Delta n$, the electron and hole bands are
separated by an energy gap ($\Delta$) because of the different
on-site electron energies on the two layers. In an applied
perpendicular $B$-field the carrier energy spectrum consists of
the four-fold degenerate Landau levels (LLs). At $E=0$ an eight
fold degenerate LL exists at energy $\epsilon=0$, the
electron-hole symmetry point. As a consequence of this degeneracy
QHSs emerge at fillings that are multiples of four, and for
$\Delta n=0$, $\sigma_{xy}$ exhibits a double $8e^2/h$ step across
$n_{tot}=0$ as shown in Fig. 2(b).

In an applied transverse $E$-field, the eight-fold degenerate LL
at $\epsilon=0$ splits into two, four-fold degenerate LLs. These
two LLs, containing electron and holes states respectively, are
symmetrically positioned with respect to $\epsilon=0$, and are
separated by the same energy gap $\Delta$ which splits the
electron and hole bands at $B=0$ \cite{mccannprl06}. In order to
get further insight into the physics of the $\nu=0$ QHS, in Fig.
2(c) we show $\rho_{xx}$ vs $\Delta n$, measured at $n_{tot}=0$
and $B=0$ T, along with the $\rho_{xx}$ vs $\Delta n$, measured at
$\nu=0$ and at $B=18$ T. These data show that as a function of
layer density imbalance, the $\rho_{xx}$ values at $\nu=0$ in high
$B$-field are comparable to the $\rho_{xx}$ measured at
$n_{tot}=0$ and $B=0$, strongly suggesting that the $\nu=0$ QHS
and the energy band-gap opening at $B=0$ as a results of layer
density imbalance have the same origin.

Next, we examine the resistivity values measured as a function of
the $B$-field, at fixed $n_{tot}$ and $\Delta n$. Figure 3 data
shows $\rho_{xx}$ and $\rho_{xy}$ vs. $B$, measured at $T=0.3$ K,
at a total density $n_{tot}=3.5\times10^{12}$ cm$^{-2}$, and for
different $\Delta n$ values. Consistent with the four-fold
degeneracy of the LLs, QHSs emerge at integer fillings that are
multiples of four, namely $\nu=4,8,12,16$ etc... Examination of
the Fig. 3 data reveals an interesting finding. The $\rho_{xx}$
measured at $\nu=8$ and $\nu=12$ is maximum at $\Delta n=0$, and
decreases with increasing $\Delta n$. This observation implies
that the $\nu=8$ and $\nu=12$ are weakest at balance ($\Delta
n=0$), and become stronger as the bilayer is imbalanced. The
dependence of the $\nu=8$ and $\nu=12$ QHS as a function of charge
imbalance is at variance with the single-particle theoretical
picture \cite{mccannprl06}, in which the LL spacing is {\it
independent} of the applied transverse field, and consequently of
$\Delta n$.

\begin{figure}
\centering
\includegraphics[scale=0.65]{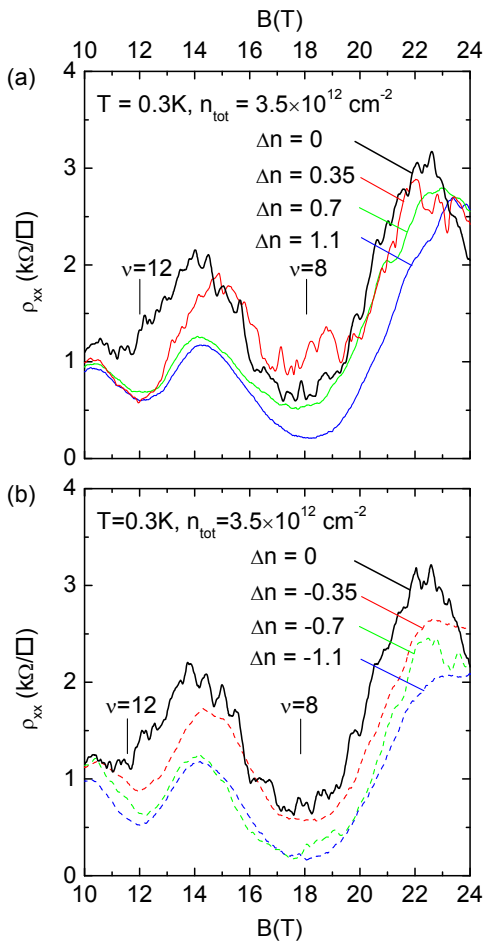}
\caption {\small{(color online) $\rho_{xx}$ vs. $B$ measured at
constant total density $n_{tot}=3.5\times10^{12}$ cm$^{-2}$, and
at different $\Delta n$ values. Panels (a) and (b) show data that
for $\Delta n\geq0$ and $\Delta n\leq0$, respectively. These data
show that the $\nu=8$ and $\nu=12$ QHSs are suppressed when the
bilayer is balanced. The $\rho_{xx}$ vs. $B$ data at $\Delta n=0$
in panels (a) and (b) are separate measurements.}}
\end{figure}

To further explore the unusual dependence on $\Delta n$ of the
QHSs $\nu=8$ and $\nu=12$, in Fig. 4 we show $\rho_{xx}$ vs. $B$
measured at $n_{tot}=3.5\times 10^{12}$ cm$^{-2}$ for different
values and signs of $\Delta n$. Figure 4(a) shows the $\rho_{xx}$
vs. $B$ data measured for $\Delta n\geq0$, when the carriers are
transferred from the top to the bottom layer, while Fig. 4(b)
shows the $\rho_{xx}$ vs. $B$ data measured for $\Delta n\leq0$,
when the carriers are transferred from the bottom to the top
layer. The data of Fig. 4 substantiate our finding that $\nu=8$
and $\nu=12$ QHSs are weakest at balance and become stronger when
the bilayer is imbalanced. A possible explanation for this finding
would be a disorder asymmetry between the two layers, which may
suppress QHS when the carriers reside predominantly in the more
disordered layer. This is however ruled out by Fig. 4 data, which
show that the $\nu=8$ and $\nu=12$ QHSs become stronger for both
positive {\it and} negative $\Delta n$, roughly symmetric with
respect to $\Delta n=0$ \cite{error}. Equally noteworthy is that
the $\rho_{xx}$ vs. $B$ measured at $\Delta n=0$ display a small
oscillation pattern, reminiscent of universal conductance
fluctuations. The $\rho_{xx}$ vs. $B$ data at $\Delta n=0$ in
panels (a) and (b) of Fig. 4 represent two different measurements,
and show that the $\rho_{xx}$ oscillations are repeatable. These
oscillations indicate edge-to-edge carrier scattering, consistent
with the presence of extended electron states at $\nu=8$ and
$\nu=12$ when these QHSs are weakest.

While the emergence of the $\nu=0$ QHS as a function of charge
imbalance can be explained by a single-particle picture as a
signature of the opening of an energy gap as a function of an
applied $E$-field \cite{mccannprl06}, the dependence of the
$\nu=8$ and $\nu=12$ QHS on $\Delta n$ cannot. Indeed, in this
picture the LLs energy spacing is independent of $\Delta n$, which
translates into QHSs that should be insensitive to an applied
transverse $E$-field. We note however that this picture does not
consider the role of disorder or electron-electron interaction.
Disorder leads to LL broadening, which may also depend on the
applied transverse $E$-field. A LL broadening which is minimum at
$\Delta n=0$ is consistent with the observed trend of Fig. 3 and
4. However, a theoretical study \cite{Ma2009} considering the role
of disorder in graphene bilayers suggests that the QHSs strength
in a disordered bilayer is independent of the applied $E$-field.
Another mechanism that can explain the $\nu=8$ and $\nu=12$ QHSs
dependence of $\Delta n$, is the electron-electron interaction in
this system. While the role of electron-electron interaction on
the LLs energy spacings in graphene bilayers has not been examined
theoretically, Min {\it et~al.} \cite{Minprb08} showed that at
$B=0$ T the exchange interaction favors the spontaneous charge
transfer between the layers, when a transverse $E$-field is
applied.

In summary we report a magneto-transport study in a dual-gated
graphene bilayer, in which the total density and charge imbalance
are independently controlled. Concomitant with the energy band-gap
opening at $B=0$ in the presence of an applied transverse
$E$-field, we observe the emerge of a QHS at filling factor
$\nu=0$, in agreement with single-particle theory
\cite{mccannprl06}. Surprisingly, the $\nu=8$ and $\nu=12$ QHSs
are suppressed when the bilayer is balanced, and become stronger
when a transverse $E$-field is applied, rendering the bilayer
imbalanced. This observation is at variance with the
single-particle picture, in which the QHS energy gaps are
independent of the $E$-field, and strongly suggests that
electron-electron interaction or disorder play a role in
stabilizing these QHSs.

We thank A. H. MacDonald, B. Sahu, and S. K. Banerjee for
discussions, and NRI-SWAN Research Center for support. This work
was performed at the National High Magnetic Field Laboratory,
which is supported by NSF (DMR-0654118), the State of Florida, and
the DOE.


\begin{thebibliography}{10}
\bibitem{error}
The balance ($\Delta n=0$) point is determined with an accuracy of
$0.2\times 10^{12}$ cm$^{-2}$.
\bibitem{filling}
The filling factor $\nu$ represents the number of occupied Landau
levels. The degeneracy of each Landau level is $eB/h$, and
$\nu=n_{tot} \cdot h/eB$; $h$ is Planck's constant.
\end{thebibliography}
\end{document}